%% file: prl.tex
\documentclass[twocolumn,showpacs,aps,prd,superscriptaddress,floatfix]{revtex4}
\usepackage{graphicx}
\usepackage{verbatim}
\usepackage{dcolumn}
\usepackage{bm}
\usepackage{amsmath}
\usepackage{epsfig}

\newcommand{\BaBarYear}       {08}
\newcommand{\BaBarNumber}     {035}
\newcommand{\SLACPubNumber} {13340}

\newcommand{\BaBarType}      {PUB} 

\input babarsym.tex

\def\bcount    {\ensuremath {383 \times 10^{6}} }

\def\btodgam {\ensuremath {b\to d\gamma} }
\def\btosgam {\ensuremath {b\to s\gamma} }

\def\Btorhoomgam {\ensuremath{B \to (\rho,\omega)\gamma}}
\def\BtoKgam  {\ensuremath{B\to K^*\gamma}}
\def\BtoXsgam  {\ensuremath{B\to X_s\gamma}}
\def\BtoXdgam  {\ensuremath{B\to X_d\gamma}}

\def\itypeone 	{\ensuremath{\Bz \to \pipi \gamma}}
\def\itypetwo	{\ensuremath{\Bp \to \pip \piz \gamma}}
\def\itypethree	{\ensuremath{\Bp \to \pipi \pip \gamma}}
\def\itypefour	{\ensuremath{\Bz \to \pipi \piz \gamma}}
\def\itypefive	{\ensuremath{\Bz \to \pipi \pipi \gamma}}
\def\itypesix	{\ensuremath{\Bp \to \pipi \pip \piz \gamma}}
\def\itypeseven	{\ensuremath{\Bp \to \pip \eta \gamma}}

\def\itypesone 	{\ensuremath{\Bz \to \Kp \pim \gamma}}
\def\itypestwo 	{\ensuremath{\Bp \to \Kp \piz \gamma}}
\def\itypesthree{\ensuremath{\Bp \to \Kp \pim \pip \gamma}}
\def\itypesfour {\ensuremath{\Bz \to \Kp \pim \piz \gamma}}
\def\itypesfive	{\ensuremath{\Bz \to \Kp \pim \pipi \gamma}}
\def\itypessix 	{\ensuremath{\Bp \to \Kp \pim \pip \piz \gamma}}
\def\itypesseven {\ensuremath{\Bp \to \Kp \eta \gamma}}

\def\figurebox#1#2#3{%
    \def\arg{#3}%
    \ifx\arg\empty
    {\hfill\vbox{\hsize#2\hrule\hbox to #2{\vrule\hfill\vbox to #1{\hsize#2\vfill}\vrule}\hrule}\hfill}%
    \else
    {\hfill\epsfbox{#3}\hfill}%
    \fi}

\begin{document}

{\pagestyle{empty}

\begin{flushleft}
\babar-\BaBarType-\BaBarYear/\BaBarNumber \\
SLAC-PUB-\SLACPubNumber\\
\end{flushleft}

\title{
  {\large \bf \boldmath
Measurement of $B \to X \gamma$ Decays and Determination of $|V_{td}/V_{ts}|$ 
}
}

\input authors_jul2008.tex

\date{\today}

\begin{abstract}
\noindent
Using a sample of 383 million \BB\ events collected by the \babar\ experiment, we measure 
sums of seven exclusive final states $B\to X_{d(s)}\gamma$, where $X_d$($X_s$) is a 
non-strange (strange) charmless
hadronic system in the mass range $0.6-1.8$\gevcc. 
After correcting for unmeasured decay modes in this mass range, 
we obtain a branching fraction for \btodgam\ of
$(7.2\pm 2.7(stat.)\pm 2.3(syst.))\times 10^{-6}$. 
Taking the ratio of $X_d$ to $X_s$ we
find $\Gamma(\btodgam)/\Gamma(\btosgam) = 0.033 \pm 0.013 (stat.) \pm 0.009 (syst.),$
from which we determine $|V_{td}/V_{ts}|=0.177\pm 0.043$.  

\end{abstract}


\maketitle

\pagestyle{plain}

The decays \btodgam\ and \btosgam\  are flavor-changing neutral current processes forbidden 
at tree level in the standard model (SM). 
The leading-order processes are one-loop electroweak 
penguin diagrams in which the top quark is the dominant virtual particle.
In the SM the inclusive rate for \btodgam\ is suppressed compared to \btosgam\ by a factor 
of $|V_{td}/V_{ts}|^2$, where $V_{td}$ and $V_{ts}$ are Cabbibo-Kobayashi-Maskawa matrix elements.  
Measurements of $|V_{td}/V_{ts}|$ using the exclusive modes 
$\Btorhoomgam$ and $\BtoKgam$~\cite{babarbellerhog} 
have theoretical uncertainties of 7\% from weak annihilation and 
hadronic form factors~\cite{BJZ}. A measurement of the inclusive decay $\btodgam$ relative to 
$\btosgam$ could determine $|V_{td}/V_{ts}|$ with reduced theoretical uncertainties 
compared to the exclusive modes~\cite{AAG}. 
In theories beyond the SM~\cite{bsm}, new virtual particles
may appear differently in the penguin loop diagrams for \btodgam\ and \btosgam\ and in the
box diagrams responsible for $B_d$ and $B_s$ mixing~\cite{PDG}, leading to measurable
differences in $|V_{td}/V_{ts}|$ extracted from these two methods.

We present measurements of the rare decays \BtoXdgam\ using seven exclusive final states
(see Table~\ref{tab:itypes}) in the hadronic mass range  
$0.6<M(X_d)<1.0\gevcc$ (which contains the $\rho$ and $\omega$ resonances), 
and in the previously unmeasured region $1.0<M(X_d)<1.8\gevcc$. 
We combine our results in the two mass regions and make corrections for 
decay modes that are not reconstructed 
to obtain an inclusive branching fraction for $b\to d\gamma$
in the mass range 0.6-1.8\gevcc. We perform a parallel analysis of \BtoXsgam\ using 
the equivalent seven modes (Table~\ref{tab:itypes}),
and determine the ratio of inclusive rates  
$\Gamma(b\to d\gamma)/\Gamma(b\to s\gamma)$
in the hadronic mass range $0.6<M(X_d)<1.8\gevcc$.

\begin{table}
\newcommand\TTT{\rule{0pt}{2.6ex}}
\centering
\caption{\label{tab:itypes} The reconstructed decay modes. Charge conjugate states 
are implied throughout this paper.}
\vspace*{2mm}
\begin{tabular}{ll}
\hline\hline
\BtoXdgam\ 	& \BtoXsgam\  \\  \hline 
\itypeone\ \TTT	&\itypesone\\
\itypetwo\ 	&\itypestwo\\
\itypethree\ 	&\itypesthree\\
\itypefour\ 	&\itypesfour \\
\itypefive\ 	&\itypesfive \\
\itypesix\ 	&\itypessix \\
\itypeseven\ 	&\itypesseven\\
\hline\hline
\end{tabular}
\end{table}

These measurements use a sample of \bcount\ \BB\ pairs collected at the \FourS\ resonance
with the \babar\ detector~\cite{babar} at the PEP-II $B$ factory.  
The high-energy photon is reconstructed from an isolated energy cluster in the
CsI(Tl) calorimeter, which has a shape consistent with a single photon, and an energy
$1.15 < E_\gamma^* < 3.5 \gev$ in the center-of-mass (CM) frame. 
We remove photons that can form a \piz($\eta$) candidate with 
another photon of energy greater than 30(250)\mev, if the 
two-photon invariant mass is in the range $105<m_{\gamma\gamma}<155\mevcc$ 
$(500<m_{\gamma\gamma}<590\mevcc)$. 

Charged pion and kaon candidates are measured in a 1.5~T magnetic field as 
tracks in a 5-layer silicon vertex detector and a 40-layer drift chamber, 
with a minimum momentum in the laboratory frame of 300\mevc. 
To differentiate pions from kaons we 
combine information from the detector of internally reflected Cherenkov light 
with the energy loss measured in the tracking system. 
At a typical pion energy of 1\gev, the pion 
selection efficiency is 85\%\ and the kaon mis-identification rate is 3\%. 	
Kaons are selected by inverting the pion selection criteria. 
We reconstruct $\piz(\eta$) candidates with momenta greater than 300\mevc\ 
from pairs of photons of minimum energy 20 \mev\ with an invariant mass 
$107<m_{\gamma\gamma}<145\mevcc$ $(470<m_{\gamma\gamma}<620\mevcc)$. 
The selected charged tracks, $\piz(\eta)$ candidates, and high-energy photons are combined 
to form \B\ meson candidates consistent with one of the seven \BtoXsgam\ or \BtoXdgam\ decay modes.
For \BtoXsgam\ decays one charged kaon is required, with all other tracks required to be pions. 
For \BtoXdgam\ decays, all tracks are required to be identified as pions. 
The charged particles are combined to form a common vertex with a vertex fit 
probability greater than 2\%.

Most of the backgrounds in this analysis arise from continuum
$e^+e^-\to q\bar{q}$ events, $q=(u,d,s,c)$, in which a
high-energy photon comes from either initial state radiation 
or the decay of a $\piz(\eta)$ meson. 
We require $R_2<0.9$ and $|\cos\theta_T|<0.8$, where $R_2$ is the 
ratio of the second to zeroth Fox-Wolfram moments~\cite{fox}, 
and $\theta_T$ is the angle between the photon and the thrust axis of the rest of the 
event (ROE) in the CM frame.
The ROE includes all the charged tracks and neutral energy in the calorimeter 
not used to reconstruct the $B$ candidate.

The quantity $\cos\theta_T$ and twelve other variables that distinguish between 
signal and continuum events are combined in a neural network (NN). These 
include the ratio $R'_2$, which is $R_2$ is calculated in the frame recoiling 
against the photon momentum, the $B$ meson  production angle $\theta_B^*$ in the CM frame
with respect to the beam axis, and five Legendre moments
of the ROE with respect to both the thrust axis of the ROE, 
and the direction of the high-energy photon. 
Differences in lepton and kaon 
production between background and \B\ decays are exploited by including five flavor-tagging 
variables applied to the ROE~\cite{babartag}. 
We optimize the NN configuration for maximal discrimination between 
signal and continuum background, which gives $50\%$ signal efficiency 
and $0.5\%$ misidentification of continuum background. 

We use the kinematic variables $\de = E^*_B - E^*_{\mbox{\scriptsize{beam}}}$ 
and  $\mes = \sqrt{ E^{*2}_{\rm beam}-{|\vec{p}}_{B}^{\;*2}|}$,
where $E^*_B$ and ${\vec{p}}_B^{\;*}$ are the CM energy and momentum of the $B$  
candidate, and $E^*_{\mbox{\scriptsize{beam}}}$ is the CM energy of one beam.
Signal events are expected to have a $\de$ distribution centered at zero 
with a resolution of about $30\mev$,  and an $\mes$ distribution centered at 
the mass of the $B$ meson with a resolution of about $3~\mevcc$. 
We consider candidates in the ranges $-0.3\gev < \de <0.2 \gev$ 
and $\mes >  5.22\gevcc$ to incorporate sidebands that allow the 
combinatorial background yields to be extracted from a fit to the data.
On average there are 1.75 candidates per event, and 
in events with multiple candidates we select the one with the 
best $\piz (\eta)$ mass, or, where there is no $\piz (\eta)$ we select the 
candidate with the best vertex fit probability.

The signal yields in the data are determined from two-dimensional 
unbinned maximum likelihood fits to the $\de$ and $\mes$ distributions
of the sums of all seven final states listed in Table \ref{tab:itypes}.
We consider the following contributions: signal, combinatorial backgrounds 
from continuum processes, $B\to X\piz/\eta$ decays, 
backgrounds from other $B$ decays, 
and cross-feed from mis-reconstructed signal $B\to X\gamma$ decays. 
The fits to the \BtoXdgam\  samples contain a component from misidentified 
\BtoXsgam\ decays, but we neglect the small \BtoXdgam\  background in 
the \BtoXsgam\ samples. 
The \B\ background yields are determined from a  
Monte Carlo (MC) simulation, whereas the continuum background is allowed to float in the fit. 

Each background contribution is modeled by a probability density function (PDF) 
that is determined from MC. 
The signal PDFs are the product of one-dimensional $\mes$ and $\de$ distributions 
determined from fits to the \BtoKgam\ data. 
For the signal cross-feed component, and the \BtoXsgam\ background 
in the \BtoXdgam\ fit, we use two-dimensional histogram PDFs to account for correlations. 
The contributions from $B\to X\piz/\eta$ 
are modeled by Gaussian peaks in both \de\ and \mes, where \de\ is displaced by $-80\mev$
due to the missing photon. The \BtoXsgam\ background in the \BtoXdgam\ sample  
also peaks with \de\ displaced by $-50\mev$ due to the kaon misidentification. 
Continuum and other non-peaking backgrounds are described by an ARGUS shape~\cite{argus} 
in \mes  and a second-order polynomial in \de.

\begin{figure}[bht]
\begin{center}
\begin{minipage}[htb]{1.65in}
\includegraphics[width=1.65in]{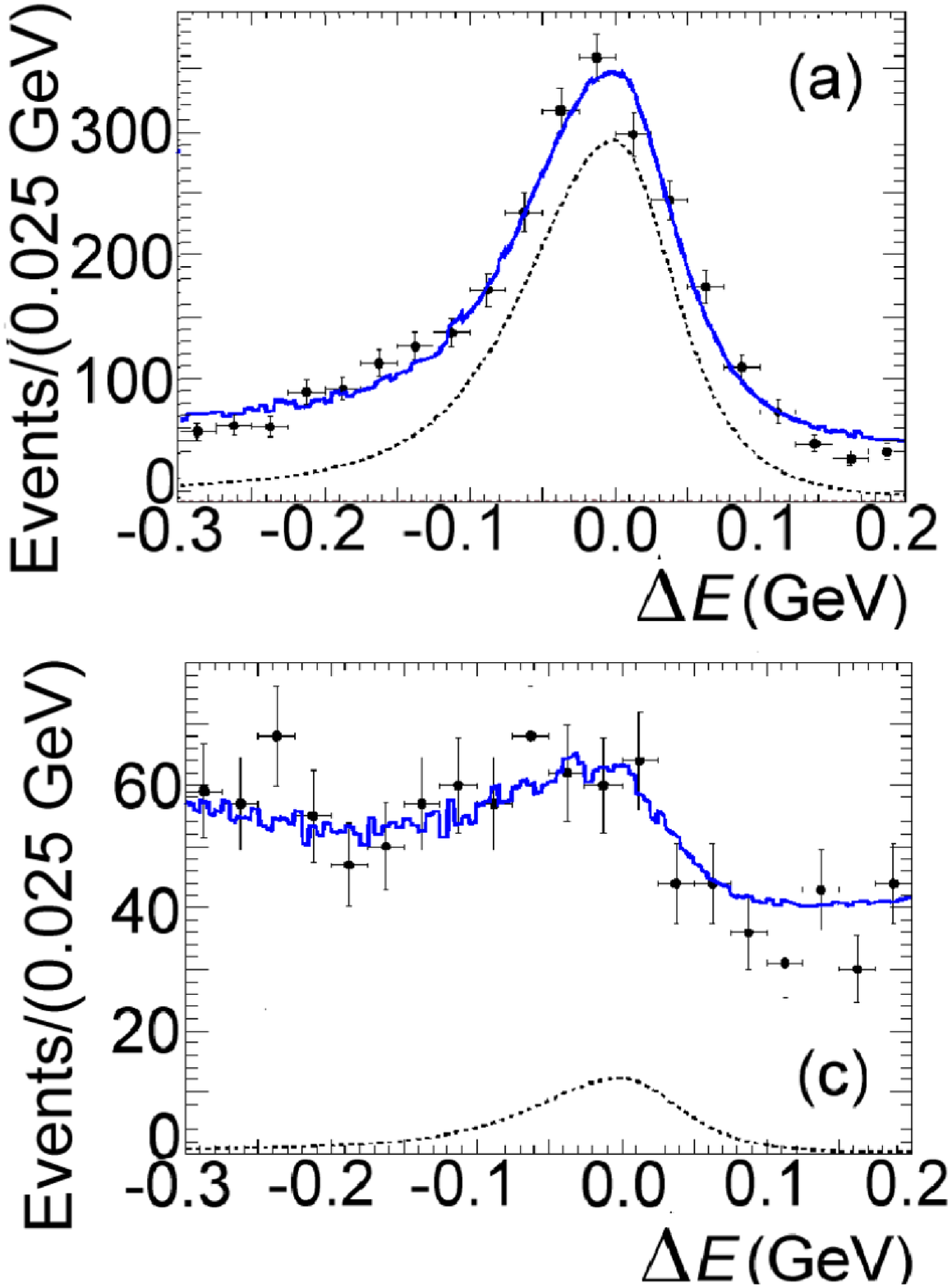}
\end{minipage}
\hfill
\begin{minipage}[htb]{1.65in}
\includegraphics[width=1.65in]{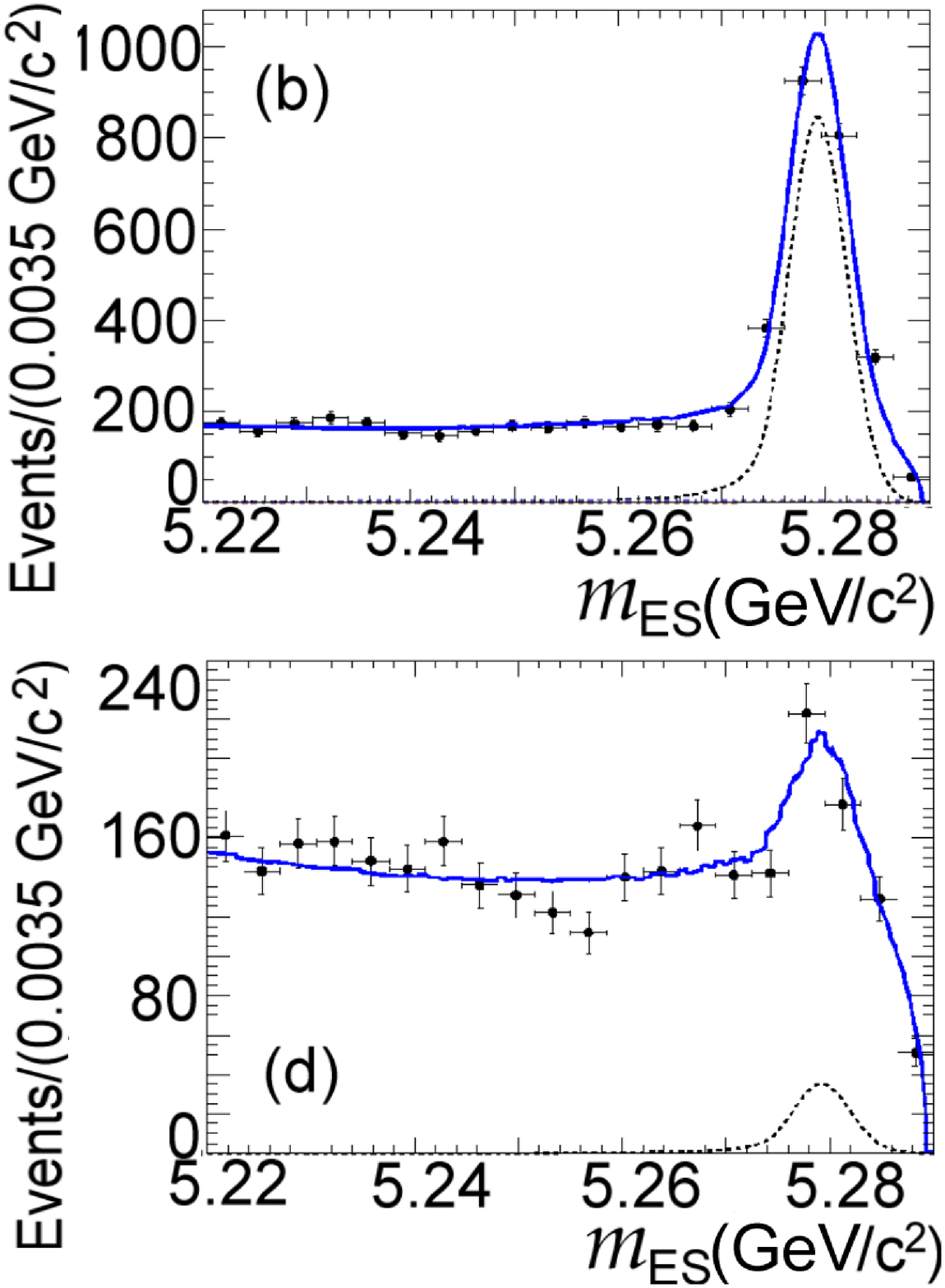}
\end{minipage}
\caption{Projections of the fits to data in the hadronic mass range 
0.6-1.0\gevcc. 
Projection of $\de$ with $5.275<\mes<5.286\gevcc$ for 
(a) \BtoXsgam\ and (c) \BtoXdgam, and 
$\mes$ with $-0.1<\de<0.05\gev$ for
(b) \BtoXsgam\ and (d) \BtoXdgam. 
Data points are compared with the sum of all the fit contributions 
(solid line) including the signal (dashed line).}
\label{fig:proj-low} 
\end{center}
\end{figure}

\begin{figure}[bht]
\begin{center}
\begin{minipage}[htb]{1.6in}
\includegraphics[width=1.6in]{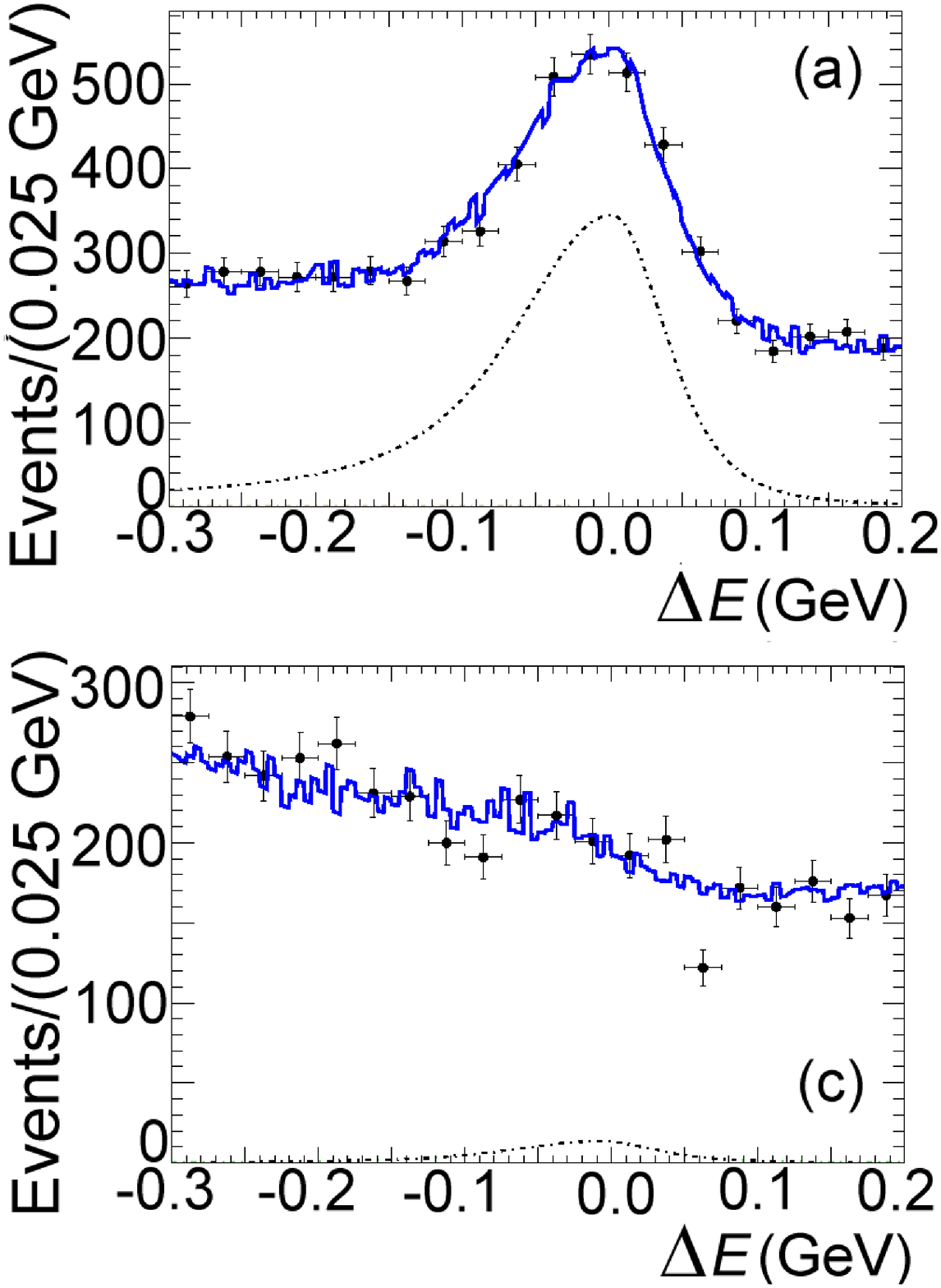}
\end{minipage}
\hfill
\begin{minipage}[htb]{1.6in}
\includegraphics[width=1.6in]{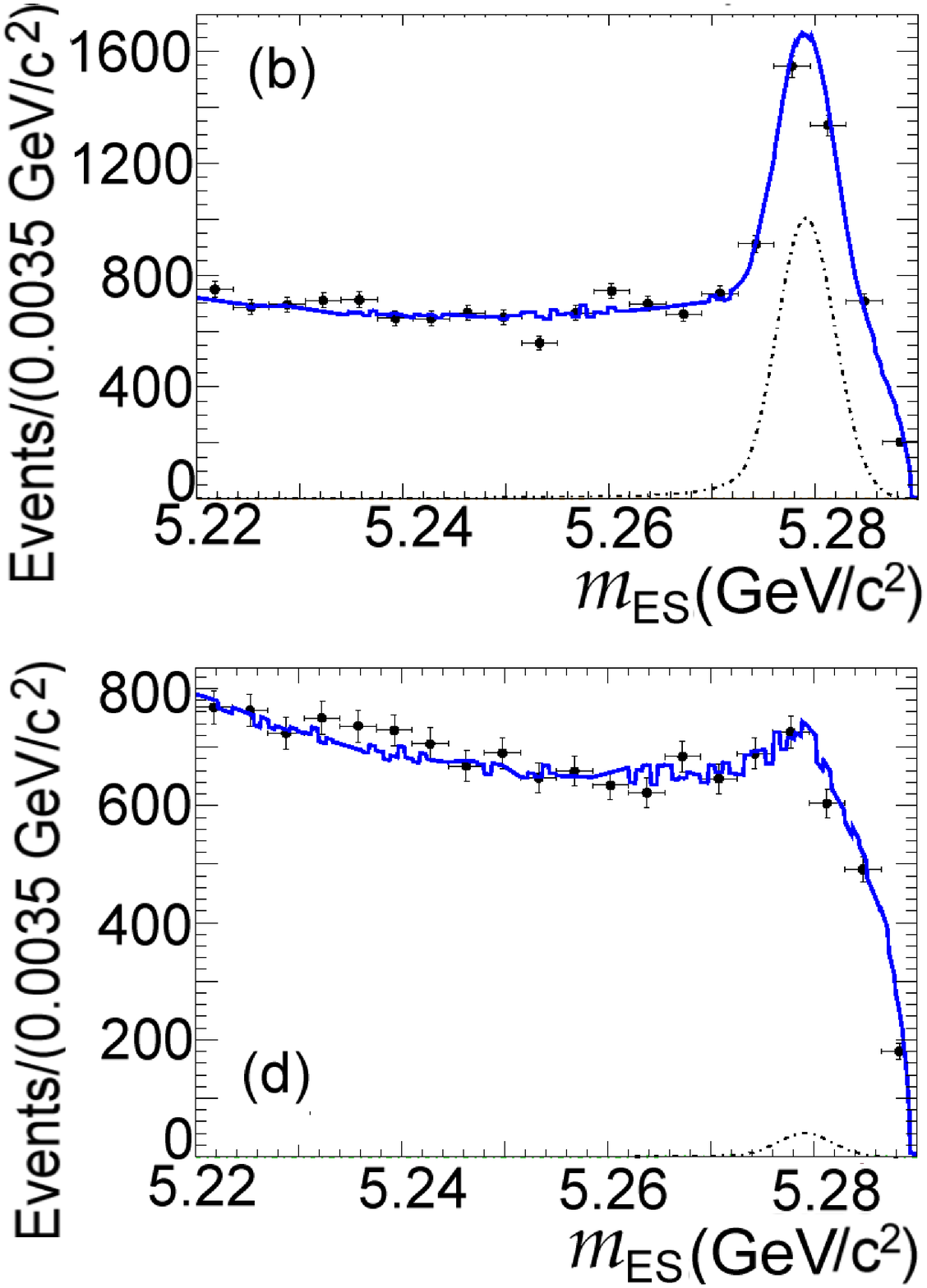}
\end{minipage}
\caption{Projections of the fits to data in the hadronic mass range 
1.0-1.8\gevcc. 
Projection of $\de$ with $5.275<\mes<5.286\gevcc$ for 
(a) \BtoXsgam\ and (c) \BtoXdgam, and 
$\mes$ with $-0.1<\de<0.05\gev$ for
(b) \BtoXsgam\ and (d) \BtoXdgam.
Data points are compared with the sum of all the fit contributions 
(solid line) including the signal (dashed line).}
\label{fig:proj} 
\end{center}
\end{figure}

We perform fits separately for \BtoXdgam\ and \BtoXsgam\ and in
the two hadronic mass ranges. 
The signal and continuum yields and the ARGUS and polynomial 
continuum shape parameters are allowed to vary.
We scale the cross-feed contribution proportionally to the fitted signal
yield, re-fit, and iterate until the fit converges. 
The fits for \BtoXsgam\ and \BtoXdgam\ are shown in the
low- and high-mass regions in Figs.~\ref{fig:proj-low}
and~\ref{fig:proj}, respectively.

The signal yields, average efficiencies and partial branching fractions for 
the sums of the seven decay modes are given in Table~\ref{tab:bfs}. 
The reconstruction efficiency depends on the distribution of 
the signal yield among the final states. For $X_s$ we measure 
the distribution of the final states in the data, but for 
$X_d$ there is no statistically useful information, so we model 
the distribution using the the phase space fragmentation model implemented in JETSET~\cite{JETSET}.

The branching fractions in Table~\ref{tab:bf2} are obtained 
after correcting for missing final states. 
The low mass \BtoXsgam\ measurement is found to be consistent
with previous measurements of the rate for \BtoKgam~\cite{hfag}, 
after accounting for the 50\% of decays to neutral kaons.
For the low mass \BtoXdgam\ region, non-reconstructed $\rho$ and $\omega$ decays 
are small and we find a 
branching fraction of $(1.2 \pm 0.5)\times 10^{-6}$, 
consistent with previous measurements of 
$\BR(\Btorhoomgam)$~\cite{babarbellerhog}. 
In the high mass region, we correct 
for missing final states with $\ge 5$ stable particles, 
or with multiple $\pi^0$s, using the fragmentation model described above.

\begin{table}
\centering
\caption{\label{tab:bfs}
Signal yields ($N_S$), average efficiencies ($\epsilon$) and 
partial branching fractions (${\cal B}$) for the measured decay modes. 
The first error is statistical, 
the second systematic.}
\vspace*{2mm}
\begin{tabular}{crcr} \hline\hline
$M(X)[\gevcc]$       & $N_S$\hspace{3 mm}	& $\epsilon$	& ${\cal B} (\times 10^{-6})\hspace{3 mm}$ 	\\ \hline
$0.6 < M(X_s) < 1.0$\   & \ $1543\pm 46$\   	& \ 8.5\%\ 	  & \ $23.7 \pm 0.7 \pm 1.7$ \\ 
$0.6 < M(X_d) < 1.0$\  	& \ $66\pm 26$\  	& \ 7.0\%\        & \ $1.2 \pm 0.5 \pm 0.1$ \\
$1.0 < M(X_s) < 1.8$\   & \ $2279\pm 75$\  	& \ 6.1\%\        & \ $48.7 \pm 1.6 \pm 4.1$ \\
$1.0 < M(X_d) < 1.8$\   & \ $107\pm 47$\  	& \ 5.2\%\ 	  & \  $2.7\pm 1.2 \pm 0.4$ \\ \hline \hline
\end{tabular}
\end{table}

\begin{table}
\centering
\caption{\label{tab:bf2}
Branching fractions ${\cal B}(\times 10^{-6})$ in the two 
hadronic mass regions $M(X)[\gevcc]$, 
after correcting for missing final states,  
and the ratios of \BR($b\to d\gamma$) to \BR($b\to s\gamma$).
The first errors are statistical, and the second are systematic, including 
the fragmentation of the hadronic system.}
\vspace*{2mm}
\begin{tabular}{cccc} \hline\hline
$M(X)$ & ${\cal B}(b\to d\gamma)$ & ${\cal B}(b\to s\gamma)$ & 
${\cal B}(b\to d\gamma)/{\cal B}(b\to s\gamma)$ \\ \hline
$0.6-1.0$	& \ $1.2\pm 0.5\pm 0.1$ & $47\pm 1\pm 3$   & $0.026\pm 0.011\pm 0.002$ \\
$1.0-1.8$	& \ $6.0\pm 2.6\pm 2.3$ & $168\pm 14\pm 33$ & $0.036\pm 0.015\pm 0.009$ \\
$0.6-1.8$       & \ $7.2\pm 2.7\pm 2.3$ & $215\pm 14\pm 33$ & $0.033\pm 0.013\pm 0.009$ \\ \hline\hline
\end{tabular}
\end{table}

\begin{table}
\centering
\caption{\label{tab:syst}
Systematic errors on the measured partial and total branching fractions ${\cal B}$. 
The final column shows systematic errors that do not cancel in the ratio 
of rates $\Gamma(b\to d\gamma)/\Gamma(b\to s\gamma)$.}
\vspace*{2mm}
\begin{tabular}{lccccc} \hline\hline
Systematic & \multicolumn{2}{c}{$M(X_s)$} & \multicolumn{2}{c}{$M(X_d)$} & $X_d/X_s$ \\
Error Source & 0.6-1.0& 1.0-1.8 & 0.6-1.0 & 1.0-1.8 & Ratio \\ \hline
Tracking & 1.7\% & 1.7\% & 1.7\% & 1.7\% & \\
High-energy photon & 2.5\% & 2.5\% & 2.5\% & 2.5\% & \\
$\pi^0/\eta$ reconstruction & 1.7\% & 1.7\% & 1.7\% & 1.7\% & \\
$\pi^0/\eta$ veto & 1.0\% & 1.0\% & 1.0\% & 1.0\% & \\
$K/\pi$ identification & 2.0\% & 2.0\% & 2.0\% & 2.0\% &  2.0\% \\
Neural network & 5.0\% & 5.0\% & 5.0\% & 5.0\% & \\
$\BB$ pair counting  & 1.1\% & 1.1\% & 1.1\% & 1.1\% & \\
Fit PDFs & 2.4\% & 3.6\% & 7.0\% & 8.3\% & 8.7\% \\
Backgrounds & 0.3\% & 0.4\% & 2.4\% & 6.1\% & 5.4\% \\
Fit bias & 0.4\% & 1.7\% & 0.4\% & 3.3\% & 3.0\% \\
Fragmentation & & 3.6\%  & & 7.7\% & 8.5\% \\ \hline
Partial ${\cal B}$ & 7.0\% & 11.4\% & 10.0\% & 14.8\% & 13.8\% \\ \hline
Missing $\ge 5$ body & & 5.6\% & & 25.8\% & 21.0\% \\ 
Other missing states & & 17.0\% & & 23.8\% & 7.1\% \\ 
Spectrum Model & & 1.8\% & & 1.6\% & \\ \hline
Total ${\cal B}$ & 7.0\% & 21.2\% & 10.0\% & 38.1\% & 26.1\% \\ \hline\hline
\end{tabular}
\end{table}

The sources of systematic uncertainties
in the measurement of the branching fractions are listed 
in Table~\ref{tab:syst}. 
These include uncertainties on track reconstruction efficiency, 
\g\ and \piz/$\eta$\ reconstruction, the \piz/$\eta$ veto, the NN selection, 
and the number of $\BB$ pairs. The 2\% uncertainty on 
correct kaon/pion particle identification, and the 20\% uncertainty on kaon misidentification, 
which is a systematic on the fixed \btosgam\ background in the \BtoXdgam\ fits, do not cancel in the ratio. 
The systematic errors associated with the variation of the fit PDFs 
also do not cancel because of the very different signal to background 
ratios in the two samples. We vary the signal PDF parameters within the range allowed 
by the fit to the $\B\to K^{*}\g$ data. The normalization of the signal cross-feed is varied 
by $\pm 30\%$, and the contribution of $B\to X\piz/\eta$ by $\pm 100\%$, in 
accordance with MC studies. 
The remaining peaking $B$ backgrounds, including the \BtoXsgam\ contribution 
in the \BtoXdgam\ sample, are varied by $\pm 20\%$.
We use simulated signal and background event samples to assign a systematic
uncertainty due to the potential for 
bias in the fit method. 

There is an additional systematic error on the efficiency  
due to the uncertainties in the measured fragmentation of the 
$X_s$ hadronic system into the seven \BtoXsgam\ final states.  
The equivalent error for \BtoXdgam\ is obtained from 
the difference between our fragmentation model applied to \BtoXdgam\ 
and the fragmentation observed in \BtoXsgam\ data.
We assume that these errors are independent and so do not cancel in 
the ratio of branching fractions. 

Table~\ref{tab:syst} also shows the systematic errors associated 
with correcting the partial branching fractions for the missing final states.   
There is no information from the data on the missing fraction of 
high multiplicity final states with $\ge 5$ stable hadrons, or 
on the missing fraction of other final states with $\ge$ 1 $\pi^0$ or $\eta$ mesons.
We vary these fractions by $\pm 50\%$ of their values from the default 
phase space fragmentation.
We motivate our choice of a $\pm50\%$ variation using signal models, 
for which we mix a combination of resonances
as 50\% fractions of \BtoXsgam\ and  
\BtoXdgam\ in the mass range $1.0-1.8\gevcc$. These 
give missing fractions close to the lower limits from the
$\pm 50\%$ variations. The missing fraction errors partially cancel in the ratio 
when the $\pm 50\%$ variations are made in the same direction for $b\to d\gamma$ 
and $b\to s\gamma$.

We take the spectral shape of the high-energy photon from~\cite{kaganneubert} 
with the kinetic parameters $(m_b,\mu_{\pi}^2) = (4.65\gevcc, -0.52\gev^2)$ 
extracted from fits to $b\to s\gamma$ and 
$b\to c\ell\nu$ data~\cite{OliverHenning}. 
We vary these shape parameters in a correlated way between 
$(m_b,\mu_{\pi}^2) = (4.60\gevcc, -0.60\gev^2)$  and  
$(m_b,\mu_{\pi}^2) = (4.70\gevcc, -0.45\gev^2)$. 
There are systematic errors on the branching fractions 
from these variations, but they are small and cancel in the ratio.
The fraction of the spectrum in the mass range 0.6-1.8\gevcc
is $(51\pm 4)\%$ for $b\to d\gamma$  and $(50\pm 4)\%$ for $b\to s\gamma$. 
We do not extrapolate the ratio of branching fractions to $M_X>1.8\gevcc$, 
so these errors, which mostly cancel in the ratio, are not included in Table~\ref{tab:syst}. 
If we make this correction, we obtain $\BR(\btodgam)=(1.4 \pm0.5 \pm 0.4 \pm 0.1)\times 10^{-5}$
and $\BR(\btosgam) =( 4.3 \pm 0.3 \pm 0.7 \pm 0.2)\times 10^{-4}$, 
where the first error is statistical, the second systematic and the third accounts 
for the uncertainty in extrapolating to the full mass range. 
The result for \BtoXsgam\ is consistent 
with the measured inclusive \btosgam\ branching 
fraction of $(3.55\pm 0.24)\times 10^{-4}$~\cite{hfag}. 

We convert the ratio of branching fractions from the full mass range 0.6-1.8\gevcc, 
$\Gamma(\btodgam)/\Gamma(\btosgam) = 0.033 \pm 0.013 \pm 0.009 $, 
into a value for  $|V_{td}/V_{ts}|$ using Table 1 and 
Equation (26) of~\cite{AAG}. 
The result is $|V_{td}/V_{ts}| = 0.177\pm 0.043\pm 0.001$,
where the first error is experimental, including systematic errors, 
and the second error is theoretical. The theoretical error includes 
uncertainties on the CKM parameters $\bar{\rho}$ and $\bar{\eta}$, 
and on $1/m_c^2$ and $1/m_b^2$ corrections, but does not include an uncertainty 
for the restriction of the measurement of the ratio to hadronic masses 
below 1.8\gevcc.   

As a check, we use the low mass region to 
determine $|V_{td}/V_{ts}|$ using predictions for exclusive \Btorhoomgam\
and \BtoKgam\ from~\cite{BJZ}.
We find $|V_{td}/V_{ts}| = 0.214\pm 0.046\pm 0.028$  
where the first error is experimental and the second is theoretical.
This is in good agreement with previously published results~\cite{babarbellerhog}. 

In summary we have made the first measurement of \BtoXdgam\ decays in 
the hadronic mass range up to 1.8\gevcc, and have extracted 
$|V_{td}/V_{ts}|$ from an inclusive model with small theoretical uncertainties. 
These results are consistent with the measurements of $|V_{td}/V_{ts}|$ from the 
exclusive decays \Btorhoomgam~\cite{babarbellerhog},
and with $B_s/B_d$ oscillations~\cite{PDG}. 

\input{acknow_PRL}

\end{document}

%% file: authors_jul2008.tex
%
\author{B.~Aubert}
\author{M.~Bona}
\author{Y.~Karyotakis}
\author{J.~P.~Lees}
\author{V.~Poireau}
\author{E.~Prencipe}
\author{X.~Prudent}
\author{V.~Tisserand}
\affiliation{Laboratoire de Physique des Particules, IN2P3/CNRS et Universit\'e de Savoie, F-74941 Annecy-Le-Vieux, France }
\author{J.~Garra~Tico}
\author{E.~Grauges}
\affiliation{Universitat de Barcelona, Facultat de Fisica, Departament ECM, E-08028 Barcelona, Spain }
% ...
\collaboration{The \babar\ Collaboration}
\noaffiliation

%% file: acknow_PRL.tex
We are grateful for the excellent luminosity and machine conditions
provided by our \pep2\ colleagues, 
and for the substantial dedicated effort from
the computing organizations that support \babar.
The collaborating institutions wish to thank 
SLAC for its support and kind hospitality. 
This work is supported by
DOE
and NSF (USA),
NSERC (Canada),
CEA and
CNRS-IN2P3
(France),
BMBF and DFG
(Germany),
INFN (Italy),
FOM (The Netherlands),
NFR (Norway),
MES (Russia),
MEC (Spain), and
STFC (United Kingdom). 
Individuals have received support from the
Marie Curie EIF (European Union) and
the A.~P.~Sloan Foundation.

%% file: prl.bbl
\begin{thebibliography}{99}

\bibitem{babarbellerhog}
B.~Aubert {\it et al.} [\babar\ Collaboration],
  Phys.\ Rev \ Lett.\ {\bf 98}, 151802 (2007);
D.~Mohapatra {\it et al.} [Belle Collaboration],
  Phys.\ Rev \ Lett.\ {\bf 96}, 221601 (2006).

\bibitem{BJZ}
P.~Ball, G.~Jones  and  R.~Zwicky, 
Phys.\ Rev.\ D \ {\bf 75}, 054004 (2007). 

\bibitem{AAG}
A.~Ali, H.~Asatrian  and  C.~Greub, 
Phys.\ Lett. \ B \ {\bf 429}, 87 (1998).

\bibitem{bsm}
S.~Bertolini, F.~Borzumati  and  A.~Masiero, \npb{294}, 321 (1987);
H.~Baer  and   M.~Brhlik, Phys.\ Rev.\ D \ {\bf 55}, 3201 (1997);
J.~Hewett  and  J.~Wells, Phys.\ Rev.\ D \ {\bf 55}, 5549 (1997);
M.~Carena {\it et al.}, Phys.\ Lett.\ B \ {\bf 499}, 141 (2001).

\bibitem{PDG} W.-M. Yao {\it et al.}, J.\ Phys. G {\bf 33}, 1 (2006). 

\bibitem{babar}
B.\ Aubert {\em et al.} [\babar\ Collaboration],
Nucl.\ Instrum.\ Methods \ A \ {\bf 479}, 1 (2002).

\bibitem{fox} 
G.~C.~Fox  and  S.~Wolfram, Nucl. \ Phys. \ B \ {\bf 149}, 413 (1979).

\bibitem{babartag}
B.~Aubert {\it et al.} [\babar\ Collaboration],
Phys.\ Rev.\ Lett.\  {\bf 89}, 201802 (2002).


\bibitem{argus} H. Albrecht {\it et al.} [ARGUS Collaboration],
Phys.\ Lett.\ B \ {\bf 185}, 218 (1987).

\bibitem{JETSET}
T. Sj¨ostrand, hep-ph/9508391;
T. Sj¨ostrand, Comput. Phys. Commun. {\bf 82}, 74 (1994).

\bibitem{hfag}
E.~Barberio {\it et al.} [Heavy Flavor Averaging Group], 
arXiv:0704.3575 (hep-ex) (2007).


\bibitem{kaganneubert}
A.~L.~Kagan  and  M.~Neubert, 
  Phys.\ Rev.\ D \ {\bf 58}, 094012 (1998).


\bibitem{OliverHenning}
O.~Buchm\"uller  and  H.~Fl\"acher, 
Phys.\ Rev.\ D \ {\bf 73}, 073008 (2006). 


\end{thebibliography}
